\documentclass[aps,prl,twocolumn,superscriptaddress,amsfont,graphicx,preprintnumbers]{revtex4-1}
\renewcommand{\section}[1]{\textbf{\textit{#1.---\!}}}

\usepackage[pdftex]{graphicx}

\usepackage[colorlinks=true
,urlcolor=blue
,anchorcolor=blue
,citecolor=blue
,filecolor=blue
,linkcolor=blue
,menucolor=blue
,pagecolor=blue
]{hyperref}
\usepackage[all]{hypcap}

\pdfoutput=1
\usepackage{bbm}
\usepackage{amsmath,amssymb}
\usepackage[usenames,dvipsnames]{color}
\usepackage{slashed} 

\makeatletter
\def\endfmffile{%
  \fmfcmd{\p@rcent\space the end.^^J%
          end.^^J%
          endinput;}%
  \if@fmfio
    \immediate\closeout\@outfmf
  \fi
  \IfFileExists{\thefmffile.mp}{\immediate\write18{mpost \thefmffile}}{}
  \let\thefmffile\relax
}
\makeatother
\usepackage[normalem]{ulem}

\newcommand {\beq} {\begin{equation}}
\newcommand {\eeq} {\end{equation}}
\newcommand {\bea} {\begin{eqnarray}}
\newcommand {\eea} {\end{eqnarray}}
\newcommand{\comment}[1]{}

\newcommand{\dhd}{d_{\rm had}}
\newcommand{\thd}{\theta_{\rm had}}

\newcommand{\ignore}[1]{}
\renewcommand{\slash}[1]{#1\!\!\!\!/}

\allowdisplaybreaks

\begin{document}
\title{Hadronic Calorimeter Shower Size: Challenges and Opportunities for Jet Substructure in the Superboosted Regime}

\author{Shikma Bressler}
\affiliation{Department of Particle Physics and Astrophysics, Weizmann Institute of Science, Rehovot 76100, Israel}
\author{Thomas Flacke}
\affiliation{Department of Physics, Korea University, 145 Anam-ro, Seongbuk-gu, Seoul 136-713, Korea}
\author{Yevgeny Kats}
\affiliation{Department of Particle Physics and Astrophysics, Weizmann Institute of Science, Rehovot 76100, Israel}
\author{Seung J. Lee}
\affiliation{Department of Physics, Korea University, 145 Anam-ro, Seongbuk-gu, Seoul 136-713, Korea}
\affiliation{School of Physics, Korea Institute for Advanced Study, Seoul 130-722, Korea}
\author{Gilad Perez}
\affiliation{Department of Particle Physics and Astrophysics, Weizmann Institute of Science, Rehovot 76100, Israel}

\begin{abstract} 
Hadrons have finite interaction size with dense material, a basic feature common to known forms of hadronic calorimeters (HCAL). 
We argue that substructure variables cannot use HCAL information to access the microscopic nature of jets much narrower than the hadronic shower size, which we call superboosted massive jets.
It implies that roughly 15\% of their transverse energy profile remains inaccessible due to the presence of long-lived neutral hadrons.
This part of the jet substructure is also subject to order-one fluctuations.
We demonstrate that the effects of the fluctuations are not reduced when a global correction to jet variables is applied. 
The above leads to fundamental limitations in the ability to extract intrinsic information from jets in the superboosted regime.
The neutral fraction of a jet is correlated with its flavor.
This leads to an interesting and possibly useful difference between superboosted $W/Z/h/t$ jets and their corresponding backgrounds. 
The QCD jets that form the background to the signal superboosted jets might also be qualitatively different in their substructure as their mass might lie at or below the Sudakov mass peak.
Finally, we introduce a set of zero-cone longitudinal jet substructure variables and show that while they carry information that might be useful in certain situations, they are not in general sensitive to the jet substructure. 
\end{abstract}

\maketitle

There are several reasons for why the physics of highly boosted massive jets is an important field in theoretical and experimental particle physics nowadays and will continue to be so in the foreseeable future.
The first reason why these jets are interesting is very practical. In the next run of the LHC, processes in which the massive Standard Model (SM) particles, namely the top, $W$, $Z$ and Higgs, are produced at large boost will become fairly common. 
The second is that in order to explore the unknown energy frontier and look for new physics characterized by mass scales higher than ever been studied before, the searches typically face very energetic jets that sometimes originate from the massive SM degrees of freedom, and discriminating them from light (QCD) jets is an important task for any future high energy collider.
The third is that in a very large class of SM extensions --- motivated by either naturalness or simply the idea of minimal flavor violation --- the expected robust signals are connected to events with boosted massive SM particles and almost nothing else~\cite{Seymour:1993mx,Butterworth:2002tt,Agashe:2006hk,Lillie:2007yh}.

Just to have a concrete discussion, suppose we would like to examine in detail whether the discovered Higgs particle completely solves the SM unitarity problem or maybe it is not a pointlike particle and thus deviations from the SM predictions are expected (see Refs.~\cite{Englert:2015oga,Azatov:2015oxa,Azatov:2014jga} for recent related discussions).
This basic test of the SM Higgs mechanism involves looking at the process $W_L W_L \to V V$ ($V=W_L/Z_L/h$, with the subscript $L$ denoting longitudinal polarization) at large invariant masses, $m_{VV}\gg m_V$.
To have a reasonable measurement of $m_{W_L W_L}\,$, at least one of the $W$'s should decay hadronically, resulting in a narrow massive jet.
Such a jet would have a typical opening angle of the order of twice its mass divided by its transverse momentum, $\theta_J\sim 2m_J/p_T$.
One might naively think that no fundamental problem arises when the boost is increased and $\theta_J$ is decreasing. 
In the future all that would be required is to improve the HCAL granularity in the transverse direction such that the sizes of the basic hadronic cells divided by their distance from the interaction point will be smaller than $\theta_J$. 
Our main point is that this conclusion is incorrect because the interactions between hadrons and any known HCAL material produce a shower with a typical transverse size, $\dhd$.
For a given material and detector architecture one can then define a minimal angular size $\thd$ below which the transverse jet substructure information in the HCAL will be washed out regardless of how fine the HCAL is made.
We call superboosted massive jets the massive jets for which the typical opening angle is smaller than the hadronic shower angular size, $\theta_J\lesssim\thd$.
Such jets will suffer from the fact that the part of the perturbative information carried by the effectively stable neutral particles, such as the neutron and $K_L$, and for high boosts also $K_S$, $\Lambda$ and $\Xi^0$, cannot be recovered, as such particles are traceless outside the HCAL.

\section{Superboosted massive jets and finite hadronic shower size}\label{sec:super}
In a typical detector, the HCAL is built up to contain all hadrons produced in an event and measure their energy and direction. 
Despite the complicated nature of the interaction of hadrons with material (for reviews, see, {\it e.g.}, Refs.~\cite{Leroy:2000mj,Fabjan:2003aq,Akchurin:2012zz}), a phenomenological description of the average longitudinal and lateral sizes of the induced showers as a function of the hadron energy is available. The hadrons relevant to our discussion are very energetic. For example, for $W$ jets with $p_T$ near 3 (10) TeV, the three leading long-lived hadrons carry on average energies of 1200 (2700), 700 (1500), 490 (1100)~GeV, and the three leading neutral ones carry 600 (1330), 210 (470), 80 (190)~GeV~\cite{Sjostrand:2014zea}. Similar numbers are obtained for QCD jets.
For hadrons in this range of hundreds of GeV, the dependence on the energy and species is rather mild~\cite{Leroy:2000mj,Adloff:2013kio,Bilki:2014bga}. The 95\% longitudinal containment of hadronic shower cascades, $L_{95\%}$, which is the average calorimeter depth within which 95\% of the hadronic cascade energy will be deposited, is described in terms of the nuclear interaction length, $\lambda_A$, as~\cite{Leroy:2000mj}
\beq
L_{95\%} \approx \left(6.2 + 0.8\ln(E/100~{\rm GeV})\right)\lambda_A \,.
\label{L95}
\eeq
The 95\% lateral containment for hadronic cascades, $d_{95\%}$, can also be expressed in terms of $\lambda_A$~\cite{Leroy:2000mj},
\beq
d_{95\%} \approx \lambda_A \,.
\label{R95}
\eeq
Smaller interaction lengths are obtained for materials with larger atomic weights, with 
$\lambda_A\approx 10$, $11$, $15$, $17$, $17$, $40$~cm for tungsten, uranium, copper, iron, lead, and aluminum respectively, while scintillator materials typically have larger interactions lengths. 
Effective interaction lengths of HCALs (composed of scintillator and stopping material) thus cannot be shorter than $\sim 10$~cm, with typical values, \textit{e.g.}\ in ATLAS and CMS, and the prototype future calorimeter CALICE~\cite{Bilki:2014bga}, being 20--30~cm.
 
One can then define a minimal scale,
\beq 
\dhd \approx d_{95\%} \,,
\label{dhd}
\eeq
below which the perturbative jet information becomes increasingly unresolvable in the HCAL due to overlap between the hadronic showers (see, e.g., Ref.~\cite{Adloff:2011ha}).
Thus, for any HCAL at a radial distance $r_{\rm HCAL}$ from the beam axis, one can define a reference angular size, $\thd$,  below which the jet substructure information is expected to get lost,
\beq
\thd \approx \frac{\dhd}{r_{\rm HCAL}}
\approx 0.1 \times \frac{\lambda_{\rm HCAL}}{20 {\,\rm cm}} \times \frac{2 {\,\rm m}}{r_{\rm HCAL}} \,.
\label{thd}
\eeq
While it seems very challenging to improve upon $\lambda_{\rm HCAL}$, it is in principle possible to decrease $\thd$ by increasing the radial distance, $r_{\rm HCAL}$.
A typical opening angle of a boosted $t$ or $W$ jet is $\theta_{t,W}=2m_{t,W}/p_T$.
Thus, assuming $\lambda_{\rm HCAL}=20$~cm, to resolve the substructure of a $3\,(10)$ TeV jet the HCAL needs to be at a distance of at least $r_{\rm HCAL} \approx 2,4 \,(6,12)$ meters from the beam pipe.
Note that it means that superboosted jets might become relevant already at the LHC, since the active inner radius of the HCAL is $2.3$~m for ATLAS and $1.8$~m for CMS.
Furthermore, hadronic showers sometimes start already in the electromagnetic calorimeter (ECAL), which has an inner radius of 1.4 (1.3)~m in ATLAS (CMS).
The calorimeter shower size may or may not be the most important limitation, since an angular size of about $0.1$ describes also the granularity of the ATLAS and CMS HCALs.
However, future colliders are expected to have much better HCAL granularities (see, e.g., Ref.~\cite{Adloff:2013kio}), so the HCAL shower size will become the leading obstacle.
While scaling up the detectors would eliminate the problem, this would be very costly, not only due to the increased HCAL volume but also due to the increased volume of the magnetic field for the muon detector.
This will likely make such a solution unrealistic.

\section{Limitations of jet substructure variables without neutrals}\label{sec:fluc}
The results obtained above lead to the conclusion that in the future the energy frontier will almost unavoidably have to deal with jets in the superboosted regime. 
In this regime, jet substructure analyses will have to rely solely on information obtained by the tracker and ECAL.
Methods using only tracker and/or ECAL information have already been explored in the literature~\cite{Schaetzel:2013vka,Larkoski:2015yqa,Spannowsky:2015eba,Katz:2010mr,Son:2012mb,Calkins:2013ega,CMS:2014joa}.
Here we take a somewhat orthogonal path and attempt to characterize the unavoidable fluctuations that arise in (practically all) jet substructure variables due to the spatially unresolvable energy depositions of the neutral hadrons.
(A note on terminology: in realistic situations, each ``PF neutral'' object of CMS~\cite{CMS:2014joa} contains energy depositions of multiple almost-collinear hadrons produced in the showering and hadronization of the same parton. This commonly includes the purely electromagnetic $\pi^0$'s. Our discussion assumes such electromagnetic depositions to be perfectly resolvable, and focuses on the long-lived neutral hadrons.)

In the following, we simulate events using {\sc Pythia~8.205}~\cite{Sjostrand:2014zea} with the default settings, interfaced with {\sc FastJet}~\cite{Cacciari:2011ma}. In a more detailed study, one would also check how the results change when varying the {\sc Pythia} settings or using a different Monte Carlo (e.g., {\sc Sherpa}~\cite{Gleisberg:2008ta}), to estimate the systematic uncertainties. However, as our goal in this Letter is not to study any particular jet substructure variable in detail, but to only use several simple variables to exemplify our points, we will stick to the default settings. We have checked, nevertheless, that changing the color reconnection model from the MPI-based original {\sc Pythia~8} scheme (the default choice) to the new more QCD-based scheme or the new gluon-move model, does not have any significant effect on the results presented below. 

In Fig.~\ref{fig:neutrals}, we show the fraction of energy carried by neutrons, $K_L$'s, as well as all other neutral hadrons that due to a large boost happen to decay farther than 2~m from the beam axis, for boosted $W$ and QCD jets with $p_T = 3$ and $10$~TeV.
\begin{figure}[tb]
\vspace{2mm}
\centering
\includegraphics[width=0.49\linewidth]{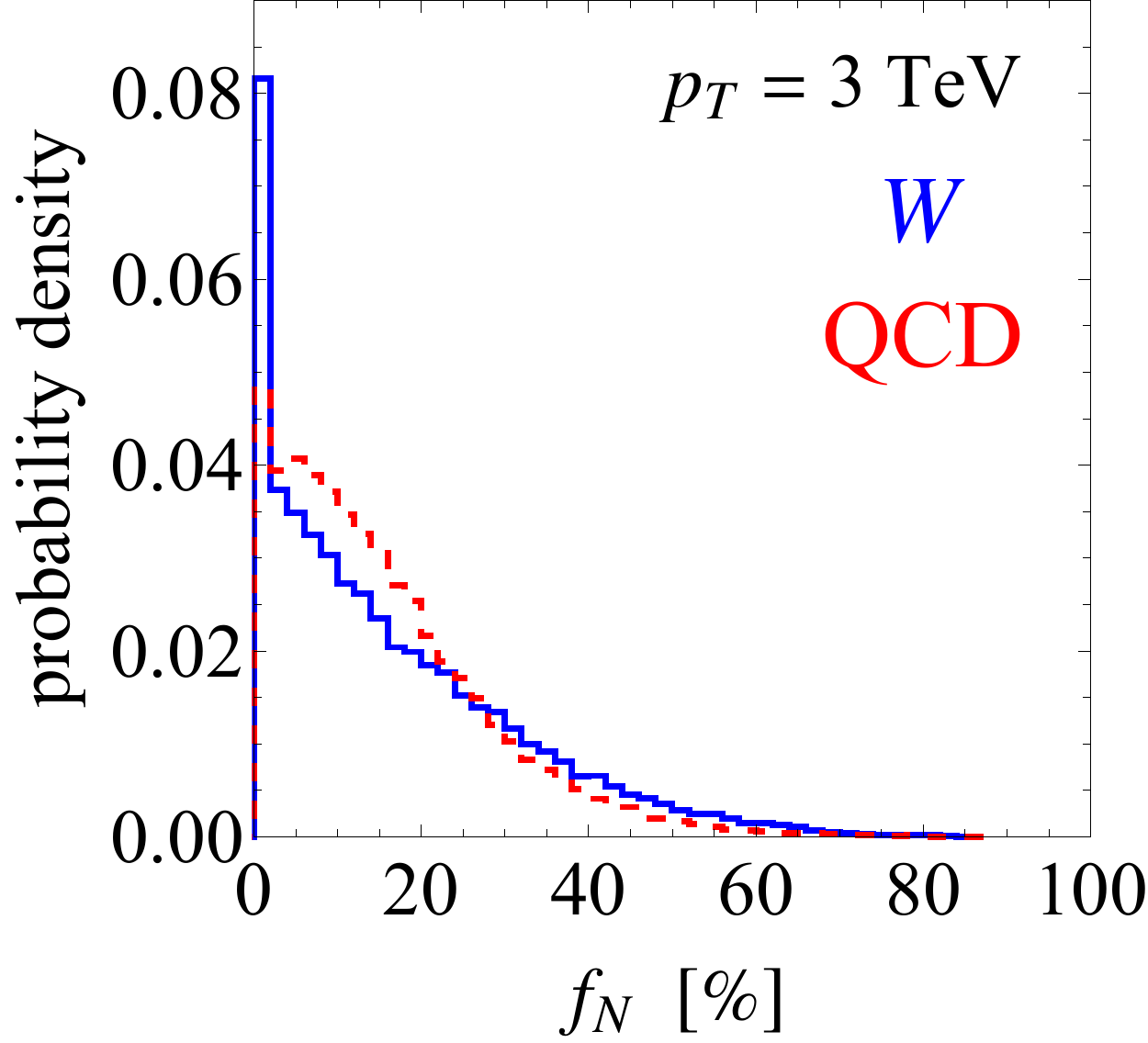}
\includegraphics[width=0.49\linewidth]{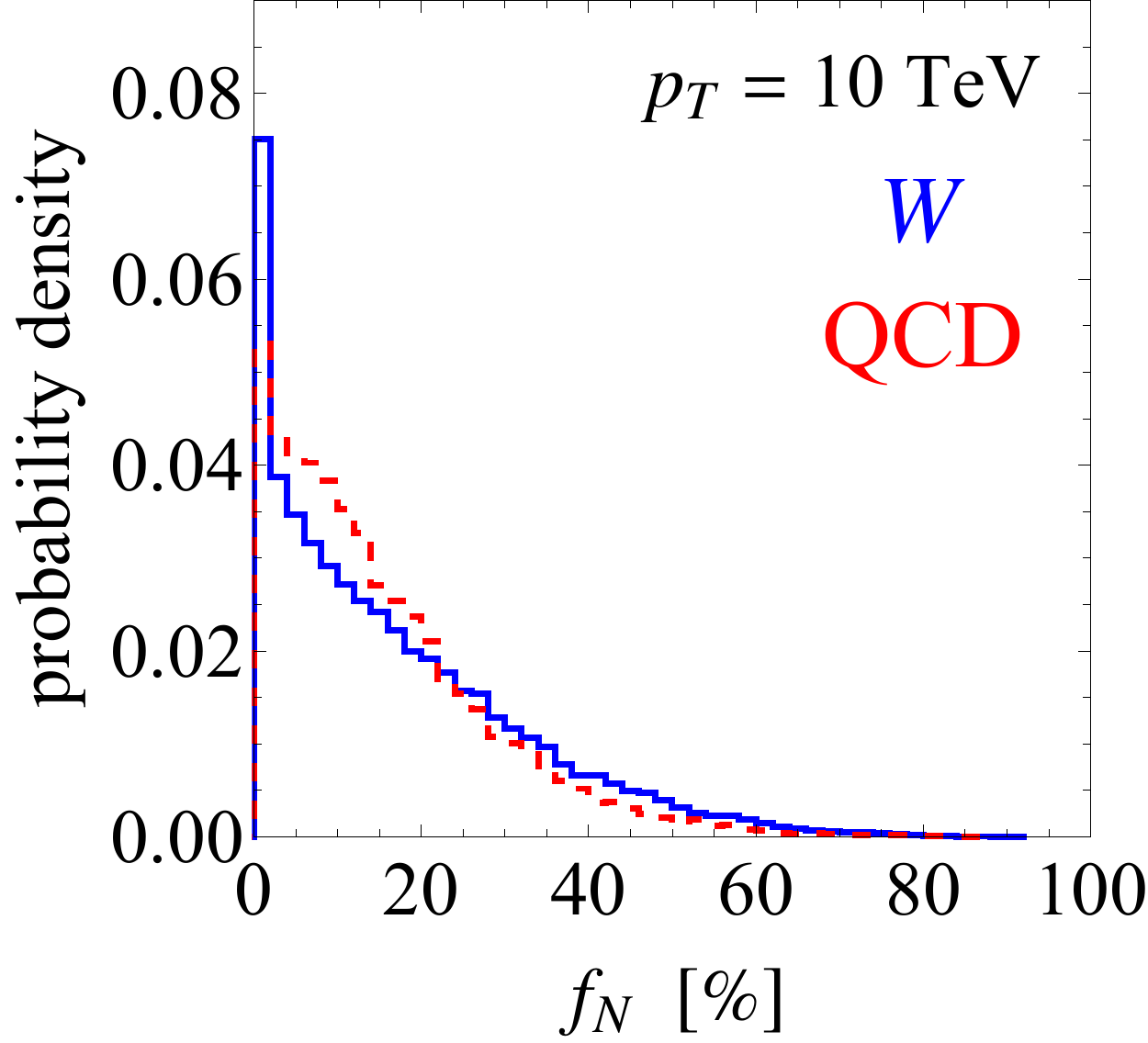}
\caption{\label{fig:neutrals}Energy fractions carried by long-lived neutral hadrons in boosted $W$ jets (solid blue) and QCD jets (dashed red) for $p_T = 3$~TeV (left) and $10$~TeV (right).}
\end{figure}
These results are based on a simulation of $WW$ and QCD events in 100~TeV $pp$ collisions.
We use as our defaults anti-$k_T$ jets~\cite{Cacciari:2008gp} with cone size $R = 3 m_W/p_T = 0.08$ ($0.024$).
Smaller cones would frequently fail to capture the $W$ decay products~\footnote{For the chosen cone sizes, the true mass of the leading jet in the $WW$ samples falls within $m_W \pm 10$~GeV in 66\% (77\%) of the cases for parton-level $p_T$ of 3 (10) TeV.}, while larger cones would increase the QCD background at $m_W$ since the average mass of a QCD jet is $\langle m_J\rangle \sim \alpha_s\, p_T R$, with the peak of the $m_J$ distribution (the Sudakov peak) being somewhat lower.
Below, we shall discuss additional impacts of using larger cones. 
The mean, $\langle f_N\rangle$, and standard deviation, $\delta f_N$, of the energy fraction taken by the neutrals in the 3 (10) TeV boosted $W$ and QCD jets are (in \%)
\beq
\hspace*{-.25cm}\langle f_N^{W,\rm QCD}\rangle = 16,15\, (17,15)\,,\ 
\delta f_N^{W,\rm QCD} = 15,13\, (15,13)\,.
\label{fN}
\eeq
It implies that tracker+ECAL based jets capture roughly $85\%\pm15 \%$ of the actual jet energy.
For subjets, obtained by reclustering the jet constituents using the anti-$k_T$ algorithm with cone size $R_{\rm subjet} = (3/4)\,m_W/p_T$, the means are similar to that of the whole jet, while the fluctuations are larger --- by factors of $1.3$--$1.4$ for each of the two leading subjets.
We note in passing that the neutral fraction depends on the flavor composition of the boosted jet partonic origin. This can potentially be used as a discriminator in certain situations. For hadronic $W$, $t$ and $h\to b\bar b$ 10~TeV jets, we find
\begin{eqnarray}
\hspace*{-.25cm} \langle f_N^{W \to c\bar s,\; W \to u\bar d}\rangle &=& 21,14, \quad \delta f_N^{W \to c\bar s,\; W \to u\bar d} = 16,14, \nonumber \\
\langle f_N^{t \to b c\bar s,\; t\to b u\bar d}\rangle &=& 18,14, \quad\, \delta f_N^{t \to b c\bar s,\; t\to b u\bar d} = 12,11, \nonumber \\
\langle f_N^{\;h \to b\bar b}\rangle &=& 17, \qquad\qquad\quad\;\, \delta f_N^{\;h \to b\bar b} = 13\,.
\end{eqnarray}

As is well known, one can apply a global rescaling to correct for the missing neutral component based on the total jet energy, $E_J$, including the energy deposited in the HCAL.
For recent discussions in the context of boosted jets, see~\cite{Schaetzel:2013vka,Larkoski:2015yqa}.
(For a formal theoretical treatment for QCD jets, see~\cite{Chang:2013rca,Chang:2013iba}.)
Jet energy resolution, which for instance in CMS is given roughly by $\sigma(E_J)/E_J \approx 1.0/\sqrt {E_J/{\rm GeV}}\oplus 0.05$~\cite{CMS-PAS-SUS-11-007}, is only a minor limitation, since already for $E_J \gtrsim 50$~GeV the associated fluctuations are below 15\%.
Now we would like to argue that such a global correction does not compensate for fluctuations in jet substructure variables. 
The reason is very simple: jet substructure, by definition, characterizes some kinematic properties of the jet's perturbative constituents, the subjets.
However, each subjet is subject to an independent fluctuation in the neutral fraction.
A global correction cannot cancel the fluctuations of the individual subjets, $f_N^i$.

Let us consider, for example, the jet mass, which is among the simplest possible jet substructure variables. 
We will show that applying a global correction to the jet does not reduce the fluctuations. 
The jet mass for boosted 2-body hadronic decays of $W/Z/h$ (signal) is dominated by just the two-prong kinematics, making it simple to describe.
For QCD jets, the mass distribution depends on the jet cone size. We shall consider two cases in the context of QCD jets as background for $W$ jets, for a fixed jet $p_T$:
(i) the $W$ mass falls in the tail region of the QCD jet mass distribution, away from the Sudakov peak, such that the two-prong approximation roughly holds (see, for instance, Refs.~\cite{Almeida:2008yp,Almeida:2008tp,Aaltonen:2011pg}) and (ii) the $W$ mass is near or below the Sudakov peak, where the QCD jet mass is partially driven by resummation of multiple emissions (see, {\it e.g.}, Refs.~\cite{Catani:1992ua,Sterman:1995fz} and references therein).

The two-prong kinematics of a narrow jet can be fully described by its energy, $E_{12}= E_1+E_2$, mass, $m_{12}^2 = E_1 E_2 \theta_{12}^2$, and the energy fraction in the softer parton/subjet, $z = E_2/E_{12} \leq \frac12$. 
Without the HCAL, one measures
\beq
m_{12,\slash N}^2 = (1-f_N^1)(1-f_N^2)\,m_{12}^2 \,,
\eeq
where the subscript $\slash N$ denotes that the neutrals are omitted.
We have neglected a possible shift in $\theta_{12}$ since the angular resolution of the tracker is very good and the subjets are very collimated.
The global jet correction accounts for the average neutral fraction by rescaling the mass according to $m_{12,{\rm corr}} = m_{12,\slash N} \times E_J/E_{J,\slash N}\,$, where $E_{J,\slash N}\,$ is the energy of all the particles in the jet that can be measured using the tracker and ECAL, namely
\beq
m_{12,\rm corr} = \frac{\sum_i E_i}{\sum_i\left(1-f_N^i\right)E_i}\; m_{12,\slash N} \;,
\label{eq:mcorr}
\eeq
where the sums are over all the subjets.
At linear order in $f_N^1$, $f_N^2$, and $y \equiv \left(\sum_i E_i - E_1 - E_2\right)/\sum_i E_i$, we obtain
\beq
\frac{m_{12,\rm corr} - m_{12}}{m_{12}} \simeq \left(\frac12 - z\right)\,(f_N^1-f_N^2) + y\, f_N^{3+} \,,
\eeq
where $f_N^{3+} \equiv \sum f_N^i E_i / \sum E_i$, with the sums in $f_N^{3+}$ starting from $i=3$.
For the mean values of $f_N^{1,2}$, the correction is perfect if we neglect the last term and the weak dependence of $\langle f_N^i\rangle$ on $E_i$.
Statistical fluctuations lead to fluctuations in $m_{12,\rm corr} - m_{12}\,$,
\begin{align}
\left(\delta\left(\frac{m_{12,\rm corr} - m_{12}}{m_{12}}\right)\right)^2 \simeq
&\; 2\left(\frac12 - z\right)^2(\delta f_N^{1,2})^2 \nonumber\\
& + \langle y\rangle^2\, (\delta f_N^{3+})^2 + \langle f_N^{3+}\rangle^2\, (\delta y)^2 .
\label{eq:fluc}
\end{align}
Note that the size of the fluctuations is $z$ dependent. 
It is interesting to see that it might be beneficial to cut away the low $z$ events as this would reduce the average fluctuation in the mass.

\begin{figure}[tb]
\vspace{1mm}
\centering
\includegraphics[width=0.49\linewidth]{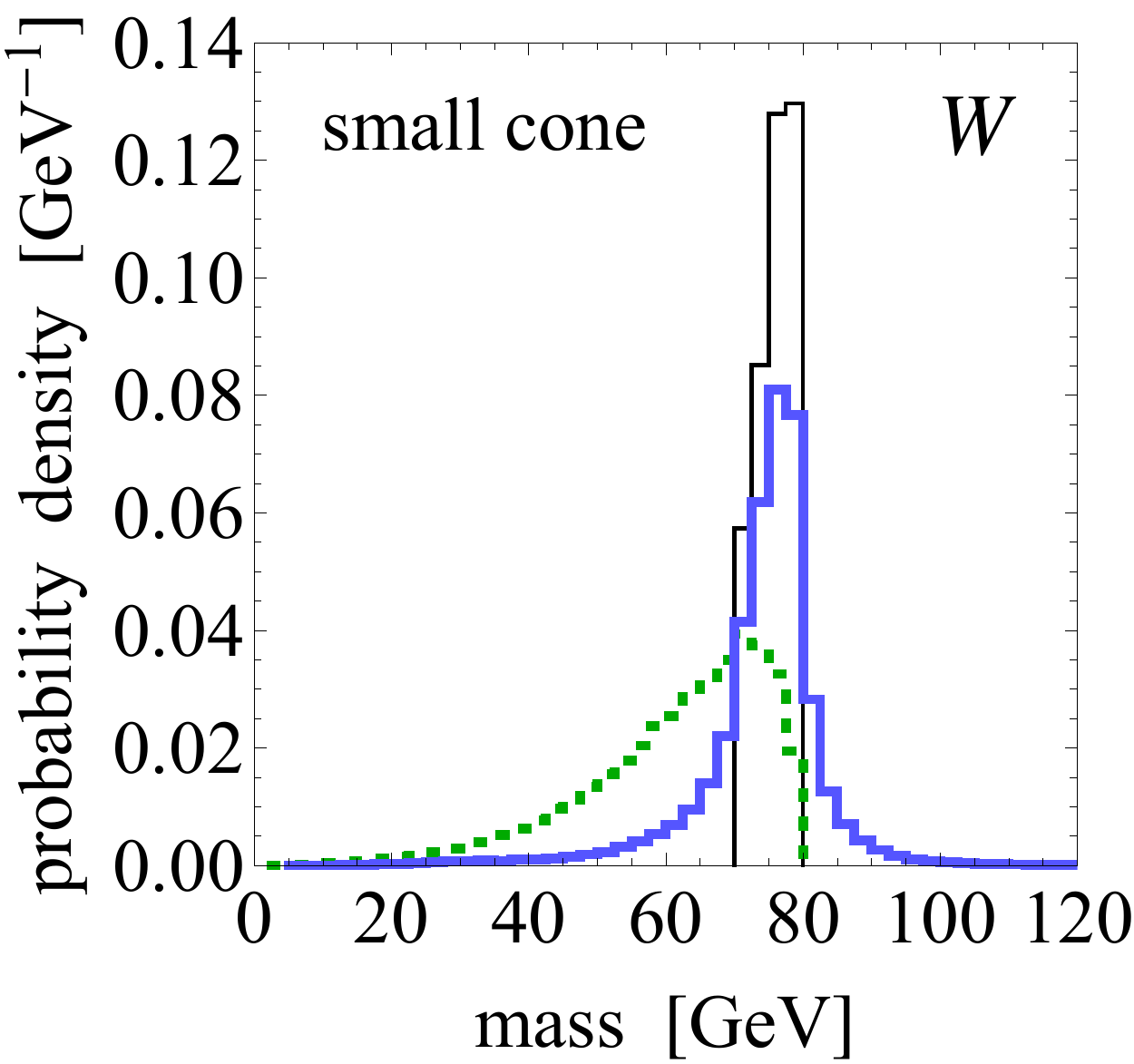}
\includegraphics[width=0.49\linewidth]{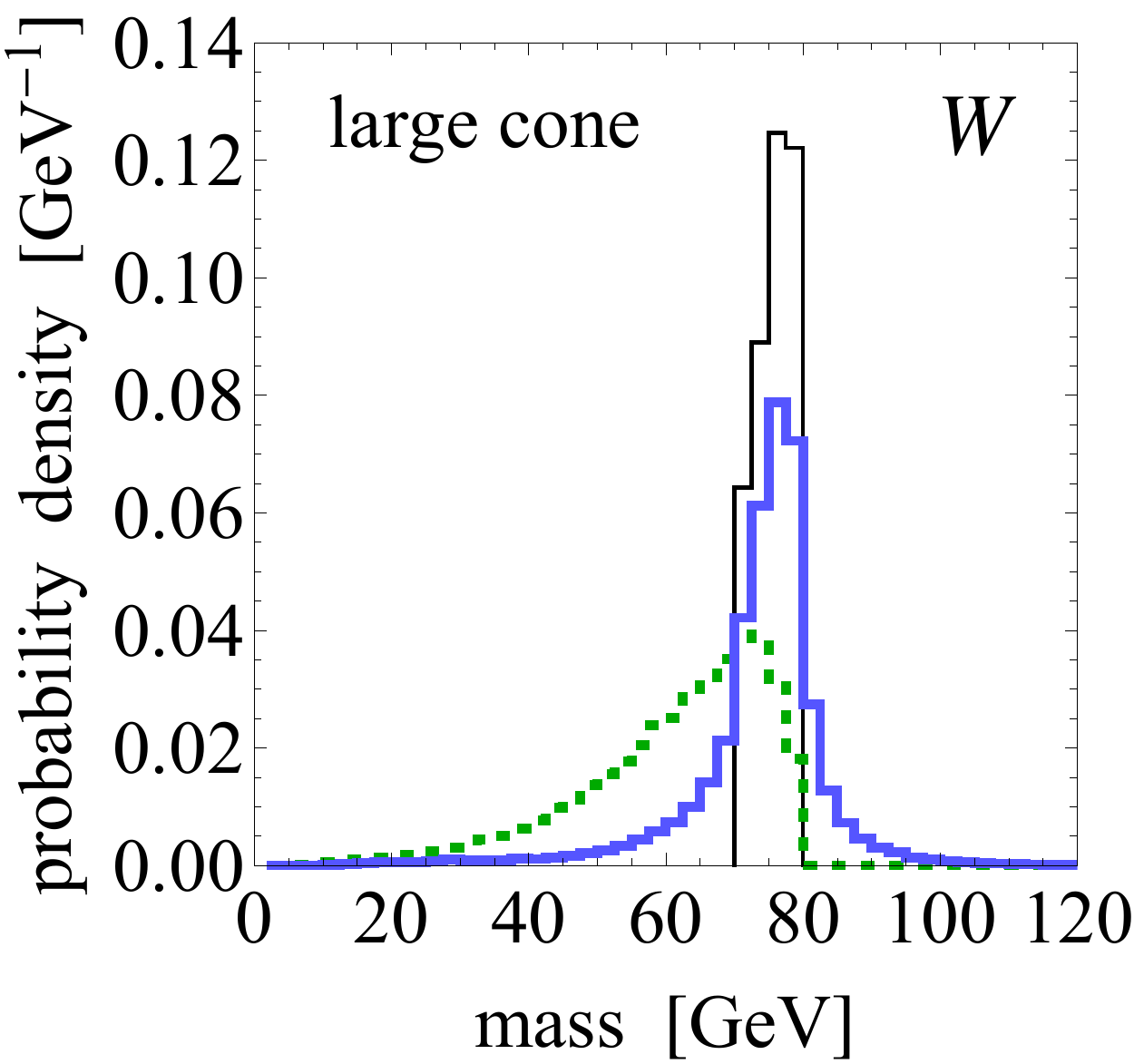}
\caption{\label{fig:mass}Jet mass based on the two leading subjets: $m_{12}$ (truth value, required to be $75 \pm 5$~GeV, thin black), $m_{12,\slash N}$ (without the neutrals, dotted green) and $m_{12,{\rm corr}}$ (corrected, thick blue) for boosted $W$ jets with $p_T = 10$~TeV, for cone sizes $R = 3 m_W/p_T$ (left) and $R = 15 m_W/p_T$ (right). In both cases $R_{\rm subjet} = (3/4)\,m_W/p_T$.}
\end{figure}

In Fig.~\ref{fig:mass} we show the distributions of the truth jet mass, $m_{12}$, the mass without the neutrals, $m_{12,\slash N}$, as well as the globally corrected one, $m_{12,{\rm corr}}$, for $W$ jets with $p_T = 10$~TeV. We focus on events where the $W$ mass is indeed captured by the two leading subjets at the truth level by requiring $m_{12}=75\pm 5$~GeV.
We see that the large fluctuations of $m_{12,\rm corr}$ relative to $m_{12}$ remain despite the correction.

\begin{figure}[tb]
\centering
\includegraphics[width=0.49\linewidth]{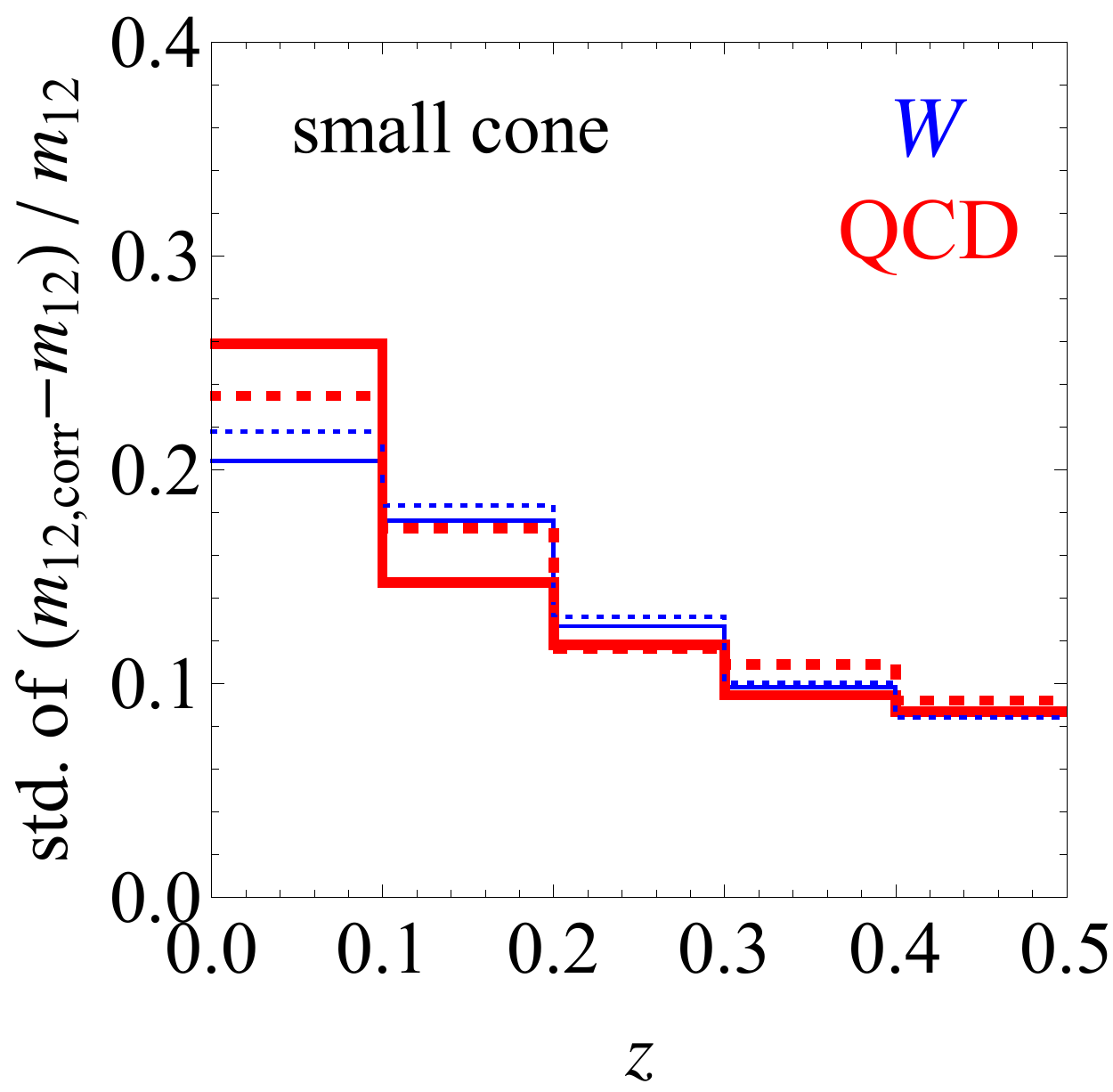}
\includegraphics[width=0.49\linewidth]{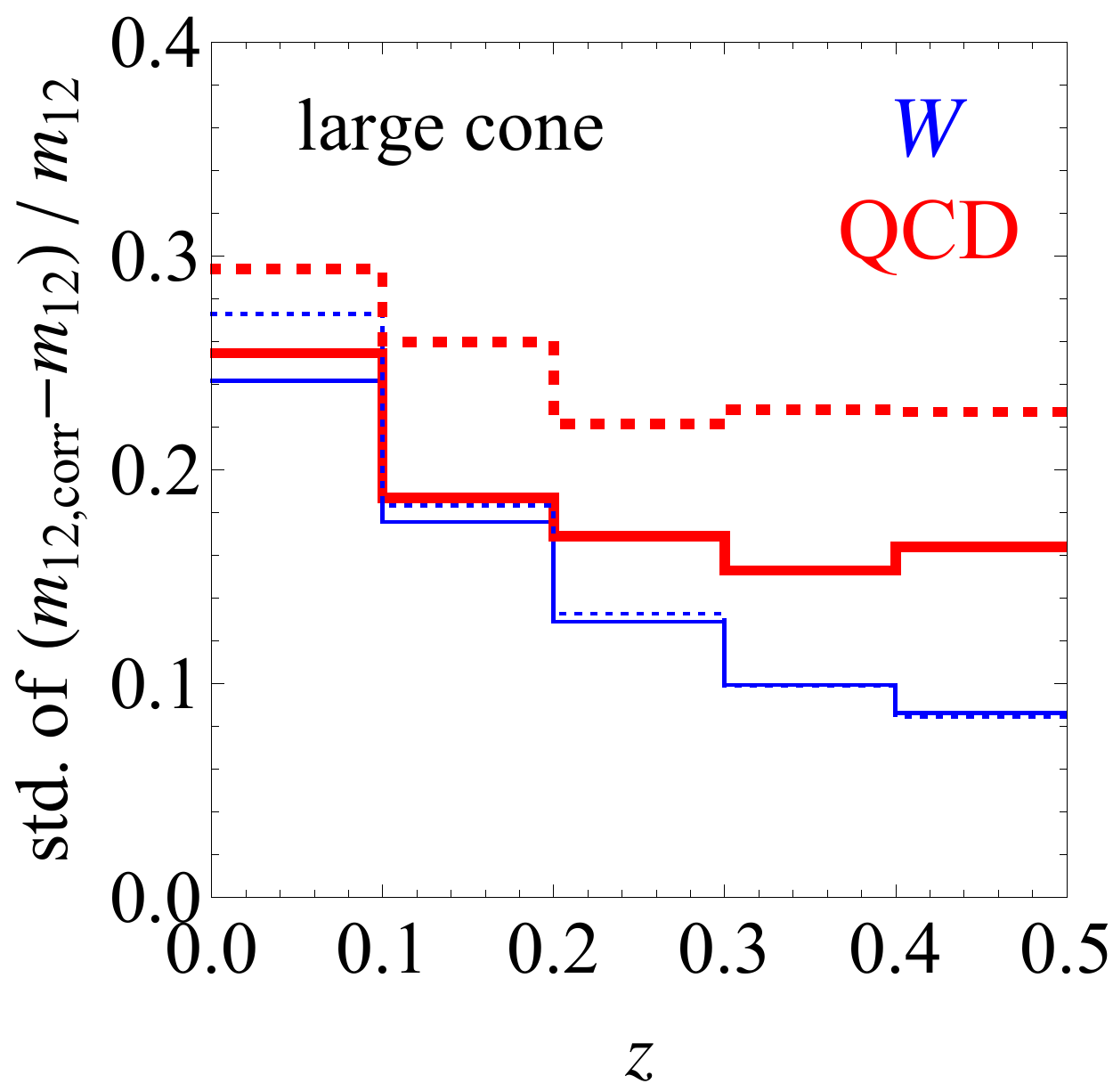}
\caption{\label{fig:mzfluc}Standard deviation of the relative offset in the jet mass after the correction (for jets with truth mass in the range $75\pm 5$~GeV) as a function of $z$, for boosted $W$ jets (thin blue) and QCD jets (thick red) with jet $p_T = 3$~TeV (solid) and $10$~TeV (dotted). On the left we use jets with cone size $R = 3 m_W/p_T$, while on the right we use large cones (the Sudakov peak case). In both cases $R_{\rm subjet} = (3/4)\,m_W/p_T$. The mean offset (not shown) is much smaller than the standard deviation.}
\end{figure}

The fluctuations are most significant for low $z$, as shown in Fig.~\ref{fig:mzfluc}, consistent with the expectation from Eq.~\eqref{eq:fluc}.
Analogous results for jets from the QCD background are also included.
For the small cone ($R = 3 m_W/p_T$), the QCD jets are far from the Sudakov peak, and are therefore dominated by the two-prong approximation and exhibit the same $z$ dependence.
Differently, when the QCD background jets are in the Sudakov peak region, which we obtain for large jet radii ($R = 9 m_W/p_T = 0.24$ for 3~TeV jets and $R = 15 m_W/p_T = 0.12$ for 10~TeV jets), they are affected by a large number of emissions, so we expect the correlation with the two-prong variable, $z$, to be rather weak.
Figure~\ref{fig:mzfluc} (right) confirms this expectation.

The only other jet-substructure variable that is independent of the mass, for two-prong kinematics, is $z$ itself.
To leading order, after fixing the mass, $W$ jets have a flat $z$ distribution while for QCD jets it is proportional to $1/z$ for small $z$'s~\cite{Almeida:2008yp}.
It is therefore possible to apply a lower cut on $z$ to enhance the signal over the corresponding QCD background~\cite{Butterworth:2008iy}, or alternatively apply an upper cut on $z$ to obtain a background-enriched sample to study massive QCD jets or have a control region.
However, the impact of the lost neutrals on the signal and background efficiencies is quite minor as the $z$ distributions of both the signal and background are pretty broad to start with.
This is also being reflected by the fact that cutting on $z$ is not particularly useful for rejecting the background.

\section{Zero-cone-size, longitudinal jet information}
Future HCALs are envisioned to have an improved granularity not only in the transverse but also in the longitudinal direction (see, \textit{e.g.}, Ref.~\cite{Adloff:2013kio}), allowing to measure the longitudinal energy deposition profile.
In principle, the profile is sensitive to the energy depositions of individual hadrons.
Separation between them is slightly aided by the fact that the shower starts at a random depth for each hadron.
The relevant so-called pion interaction length is comparable to $\lambda_A$~\cite{Adloff:2013kio}.
Remarkably, the longitudinal information is available even if the hadrons are completely collinear, when the conventional jet substructure variables, all of which depend on transverse separation, are powerless.

In practice, extracting individual contributions from a measured profile may be challenging, as there will still be a significant degree of shower overlap, the shower shapes vary significantly event-by-event~\cite{Akchurin:2012zz,Green:1993ht}, and the granularity will still be a limiting factor.
We will not analyze this in detail, but discuss how information obtained in this way can potentially be useful.

If each hard parton produced one hard hadron and a few softer ones, the longitudinal profile of a boosted $W$ jet, for example, would typically contain two relatively large humps, while a QCD jet would lead to a single and more energetic one.
That would likely be easy to see.
In practice, each high-$p_T$ parton produces several comparably energetic hadrons, so the picture is more complicated, but one might still hope that some information about the underlying partonic structure remains.
One could imagine variables such as the $p_T$ fraction carried by the leading hadron, or the number of hadrons one needs to sum to account for a certain fraction of the jet $p_T$.
If one of the hadrons is a $\pi^0(\to\gamma\gamma)$ and thus deposits all of its energy in the ECAL, it can be accounted for in a trivial way and only make the interpretation of the HCAL profile easier.

\begin{figure}[tb]
\vspace{2mm}
\centering
\includegraphics[width=0.48\linewidth]{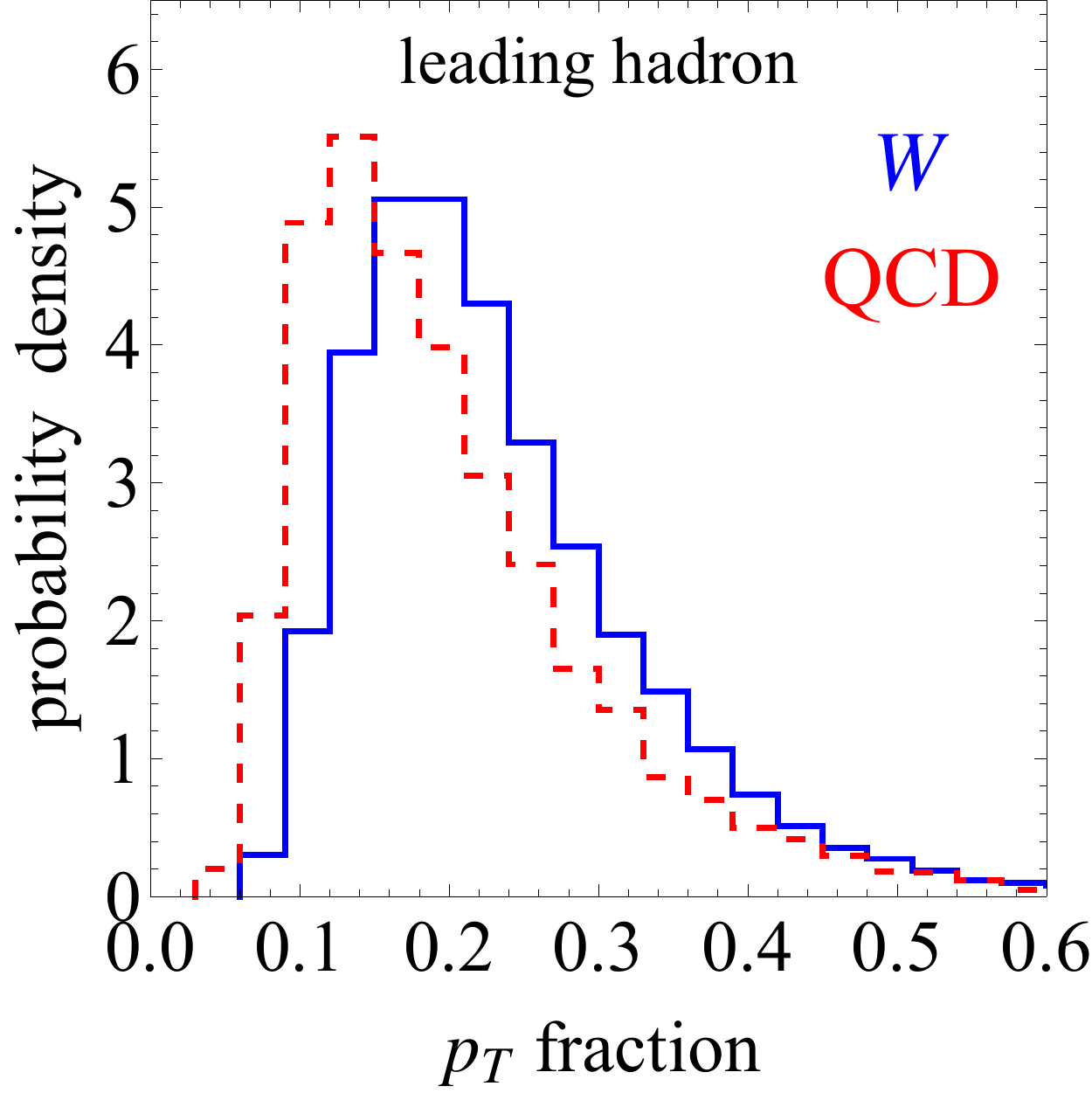}
\includegraphics[width=0.50\linewidth]{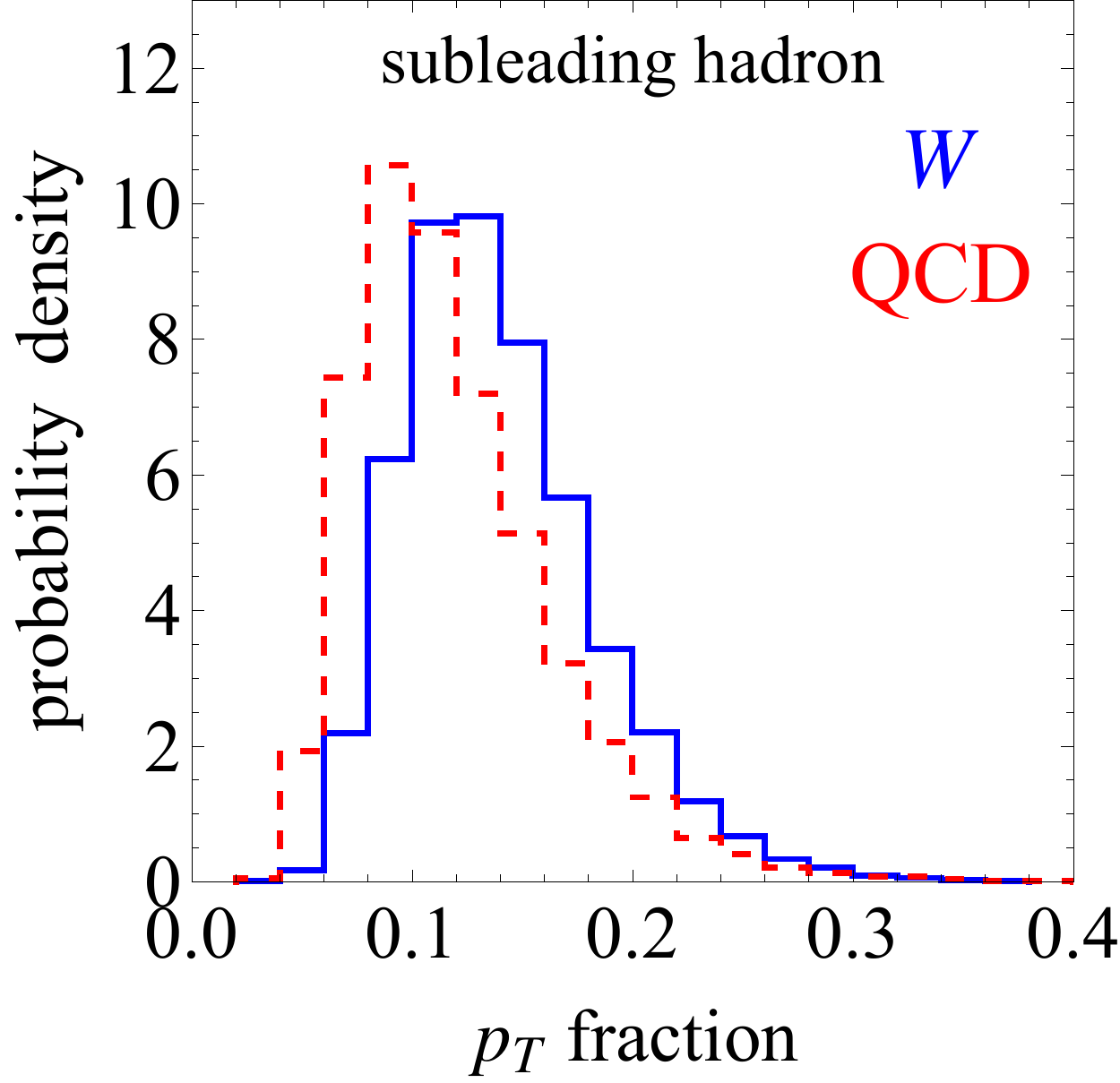}
\caption{\label{fig:pTfracs}Jet $p_T$ fractions carried by the leading hadron (left) and the second leading hadron (right) for boosted $W$ jets (solid blue) \emph{vs.} QCD jets (dashed red) for jet $p_T = 10$~TeV.}
\end{figure}

One might hope that the availability of the longitudinal profile makes the loss of transverse information less severe of an issue.
However, we find that the longitudinal information, even at the truth level (\textit{i.e.}, before simulating the HCAL showers) is quite limited, for the scenarios we analyzed in this paper.
For example, Fig.~\ref{fig:pTfracs} shows that the distributions of the $p_T$ fractions of the two leading hadrons are quite similar for boosted $W$ jets and QCD jets.
We believe that the similarity is to a large extent accidental and there may exist other scenarios in which the longitudinal variables would   be effective.
One may also consider using such variables in other contexts, \textit{e.g.}, for distinguishing jets initiated by a quark in a certain process from those initiated by a gluon in another.
It is clear though that they do not provide a general solution for the superboost regime.

\section{Overview}\label{sec:overview}
When a hadron propagates in a material, it produces a shower of a finite size.
This sets a lower bound on the angular scale  
that can be probed using HCALs in typical experiments.
We defined superboosted jets as jets originating from energetic massive particles, with an opening angle smaller than this minimal angular scale.
HCALs are thus insensitive to substructure of superboosted jets, so the information carried by the effectively-stable neutral particles is unrecoverable. 
For simplicity, we have focused on two-prong variables, the jet mass and splitting fraction.
We have demonstrated that fluctuations in the energy carried by the neutral particles lead to a smearing of the resulting jet-substructure distribution.
This qualitative feature is expected to be shared by more complicated substructure variables, as long as the signal distribution is narrow, where the fluctuations in the neutral energy fractions of the third and further subjets will play an important role.

The superboosted regime is characterized by jets with large transverse momenta.
In that regime, depending on the size of the jets, the average mass of QCD jets can be either much smaller or larger than the mass of the signal jets produced from hadronic decays of, say, the heavy SM particles, $W/Z/h/t$. 
The latter case is particularly interesting as the QCD jets and the ones originating from signal events would behave in a qualitatively different way. 
While the QCD jets will have a rich internal structure, $W/Z/h$ (and possibly $t$) superboosted jets will be much sparser.
This is somewhat similar to the difference between QCD and hadronic $\tau$ jets measured presently at the LHC.
The fundamental difference in the nature of the signal and background jets may turn out useful for suppressing the QCD background when searching for superboosted massive jets, an idea that merits a dedicated study. 

Finally, we have shown that the fraction of neutrals has a sizable flavor dependence, especially evident when comparing the decay of a boosted $W$ to second-generation quarks with its decay to first-generation quarks or with a QCD jet. 
This could be used as another handle to distinguish signal superboosted jets from their corresponding QCD backgrounds.
Distributions of the neutral fractions for various jet flavors can in fact be measured in $t\bar t$ events, with and without charm tagging of jets from $W$ decays.

\bigskip
\acknowledgments

We thank Philip Harris, David Kosower, Shmuel Nussinov and Gavin Salam for useful discussions.
We also thank Mihailo Backovic and Yotam Soreq for comments on the draft.
This research is supported in part by the I-CORE Program of the Planning and Budgeting Committee and the Israel Science Foundation (grant No.~1937/12).
GP is supported by the BSF, IRG, ISF, and ERC-2013-CoG grant (TOPCHARM \#614794).
SL and TF are supported by
the National Research Foundation of Korea (NRF) grant funded by the Korea government (MEST) (No. 2015R1A2A1A15052408),
the Basic Science Research Program through the NRF funded by the Ministry of Education, Science and Technology (No. 2013R1A1A1062597)
and the Korea-ERC researcher visiting program through the NRF (No. 2014K2a7B044399 and No. 2014K2a7A1044408).

\bibliographystyle{utphys}
\bibliography{paperbib}

\end{document}